\documentclass[aps,prb,twocolumn,showpacs,floatfix]{revtex4}

\usepackage{graphicx}
\begin{document}

\title{Properties of the Ideal Ginzburg-Landau Vortex Lattice}

\author{Ernst Helmut Brandt}
\affiliation{Max-Planck-Institut f\"ur Metallforschung,
   D-70506 Stuttgart, Germany}
\date{\today}

\begin{abstract}
The magnetization curves $M(H)$ for ideal type-II superconductors
and the maximum, minimum, and saddle point magnetic fields of
the vortex lattice
are calculated from Ginzburg-Landau theory for the entire ranges
of applied magnetic fields $H_{c1} \le H \le H_{c2}$ or induction
$0 \le B \le \mu_0 H_{c2}$ and Ginzburg-Landau parameters
$2^{-1/2} \le \kappa \le 1000$. Results for the triangular and
square flux-line lattices are compared with the results of the
circular cell approximation. The exact magnetic field $B(x,y)$
and magnetization $M(H,\kappa)$ are compared with often used
approximate expressions, some of which deviate considerably or
have limited validity. Useful limiting expressions and analytical
interpolation formulas are presented.
\end{abstract}

\pacs{74.25.Qt, 74.25.Ha, 74.20.De}

\maketitle

\section{Introduction}

   Since Abrikosov's \cite{1} prediction of the flux-line lattice
in Type-II superconductors from Ginzburg-Landau (GL) theory
\cite{2}, several approximate formulas for the magnetization
$M = B/\mu_0 - H$versus the applied magnetic field $H$ or
average induction $B$ have been published \cite{1,3,4,5,6,7}.
In these papers and below, the basic situation is considered where
a macroscopically large, homogeneous and isotropic,
long superconductor is exposed to a uniform parallel field $H$.
In this ideal case demagnetization effects, flux-line pinning,
and surface effects may be disregarded, and thus the
flux-lines are straight lines forming an ideal periodic lattice.
These results are easily extended to anisotropic superconductors
(where an anisotropic effective-mass tensor is introduced into
the GL theory) by defining an effective GL parameter
$\tilde \kappa$ that depends on the orientation of the
flux lines; this transformation works when $H$ is along a
principle symmetry axis \cite{8,9,10}. Generalization to
geometries where demagnetization effects occur, are possible
by introduction of a demagnetizing factor; but this concept
works only for homogeneous specimens with the shape of an
ellipsoid. In this case the flux lines in the bulk are still
straight and form an ideal flux-line lattice (FLL). For
other specimen shapes the FLL is distorted, i.e., the
orientation and density of the FLL varies spatially
and can be calculated only numerically \cite{11,12}.

   The aim of the present paper is to compare the widely
used approximate expressions for $M(H,\kappa)$  with the
exact value obtained numerically and to give useful
general analytic interpolation formulas valid in the entire
ranges of $H$ and $\kappa$ where the FLL exists, namely,
$H_{c1} \le H \le H_{c2}$ for $H$,  or
$0 \le B \le B_{c2} = \mu_0 H_{c2}$, and
$1/\sqrt{2} \le \kappa < \infty$ for $\kappa$,
where $H_{c1}(T)$ and $H_{c2}(T)$ are the lower and upper
critical fields and $\kappa$ is the GL parameter.
 Interestingly, such general formulas
have not been published yet, and thus the accuracy of the
commonly used expressions is not known, probably due to
the difficulty of the numerical solution of the
complex-valued GL equations. Early numerics \cite{13} used
the circular cell method (CCM), which approximates the
hexagonal unit cell of the triangular FLL (or the quadratic
unit cell of the square FLL) by a circle and the
two-dimensional (2D) solution by the 1D rotationally
symmetric solution inside this circular cell; both the
GL function and the magnetic field are forced to have
vanishing slope on this circular boundary, as the
exact solution has on the boundary of the Wigner-Seitz
cell. This method yields the exact $H_{c1}$ and is expected
to be best at low inductions $B \ll B_{c2}$ where the
flux lines are well separated. But surprisingly, the
circular cell approximation gives very good magnetization
curves at all $B$ (see Fig.~1) and even
yields the exact value of the upper critical field $H_{c2}$.
Some more exact results of the CCM are listed below.
Another method \cite{7} uses a similar circular symmetric
GL order parameter and a linear superposition of
circular symmetric magnetic fields to obtain excellent
approximate $M(H,\kappa)$, see also Ref.~\onlinecite{14}.

   An in principle exact numerical method \cite{15}
uses periodic real trial functions for the squared
GL function $|\psi(x,y)|^2$ and magnetic field $B(x,y)$
and minimizes the resulting free energy functional with
respect to a finite number of Fourier coefficients. The
same method was later applied \cite{16} to solve the
microscopic BCS-Gor'kov theory for the properties of
the FLL in the entire temperature interval
$0 \le T \le T_c$ where $T_c$ is the superconducting
transition temperature (GL theory, strictly spoken, applies
only close to $T_c$).  Recently this variational method
was improved \cite{17} by keeping the same periodic trial
functions but now solving the GL equations iteratively;
this iteration works much faster and allows to use many
more Fourier coefficients (many thousands instead of only
five in Ref.~\onlinecite{15}). I shall use this 2D iterative
precision method of Ref.~\onlinecite{17} for the
calculation of the FLL at $B > 0$.
At low inductions $B \ll B_{c2}$ this 2D method is
supplemented by an iterative circular cell method
presented in Appendix A. This 1D method yields accurate
values of $h_{c1}(\kappa) = H_{c1}/H_{c2}$, which then can
be used in interpolation formulas.
For convenience, I introduce the reduced fields
$b=B/B_{c2}$, $h=H/H_{c2}$, $m=M/H_{c2}$, such that one has
$m = b - h$,  $h_{c1} \le h \le 1$, $0 \le b \le 1$, and
$-h_{c1} \le m \le 0$.

   For completeness it should be mentioned that the isolated
vortex \cite{18} and the FLL \cite{19} have also been computed
from BCS theory (valid at all temperatures) using the
quasiclassical Eilenberger theory based on energy-integrated
Green functions. This method was recently extended to compute
the FLL structure and local density of states for s-wave
\cite{19,20,21}, d-wave \cite{21,22}, and chiral p-wave
\cite{23} superconductors. Very recently the GL method
\cite{17} was generalized phenomenologically to lower
temperatures and to charged vortices \cite{24}.
 \begin{figure}  
\includegraphics[scale=.52]{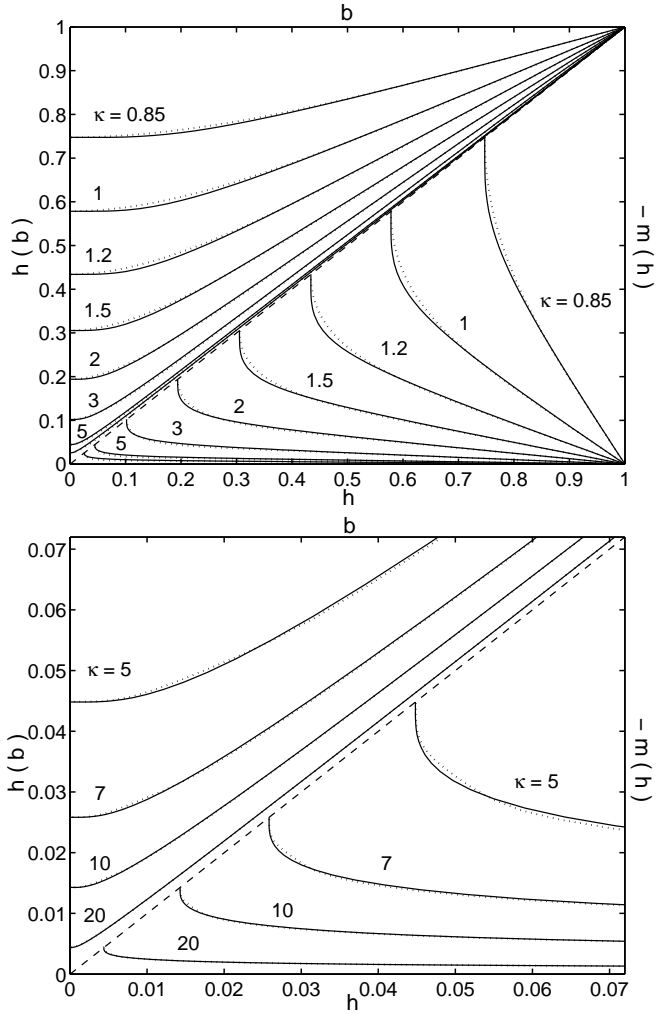}
\caption{\label{fig1} Magnetization curves of the triangular
FLL, which coincide within line thickness with the
results for the square FLL and for the FLL obtained from the
circular cell approximation, see Fig.~3 for the difference.
Shown are $h = H/H_{c2}$ versus $b=B/B_{c2}$ (upper left
triangle) and $-m = -M/H_{c2}$ versus $h$ (lower right triangle).
One has $m=b-h$. The lower panel shows an enlarged scale.
The solid lines show the exact numerical result. The dotted
lines show the simple interpolation Eq.~(22) good for
$\kappa \le 5$ (upper panel), and the combined low and high
field limit Eq.~(23)  good for $\kappa\ge 1$ (lower panel).
  }
\end{figure}
\section{Triangular and square flux-line lattices   
         and the circular cell method}

   The properties of the FLL within GL theory are calculated
by minimizing the GL free energy of the superconductor with
respect to the complex GL function $\psi({\bf r})$ and to the
vector potential ${\bf A}({\bf r})$ of the local magnetic
field ${\bf B}({\bf r}) = \nabla \times {\bf A}$.
In the usual reduced units \cite{1,2,3,4,5,6,7}  (length
$\lambda$, magnetic field $\sqrt2 H_c$, energy density
$\mu_0 H_c^2$, where $H_c$ is the thermodynamic critical field)
the spatially averaged free energy $F$ of the GL theory
referred to the Meissner state ($\psi = 1$, ${\bf B}=0$) reads
   \begin{equation} \label{1}
   F = \left\langle {(1-|\psi|^2)^2 \over 2} + \left| \left(
   {\nabla \over i\kappa} - {\bf A} \right) \psi \right|^2
   + {\bf B}^2 \right\rangle .
   \end{equation}
Here $\langle \dots \rangle = (1/V) \int d^3r \dots $ means
spatial averaging over the superconductor of volume $V$.
Introducing the supervelocity
${\bf Q}({\bf r}) = {\bf A} -\nabla\varphi/\kappa$ and the
magnitude $f({\bf r}) = |\psi|$ of
$\psi({\bf r}) = f({\bf r}) \exp[ i\varphi({\bf r}) ]$ one may
express $F$ as a functional of the real and gauge invariant
functions $f$ and ${\bf Q}$,
   \begin{equation} \label{2}
   F = \left\langle {(1-f^2)^2 \over 2} + {(\nabla f)^2 \over
   \kappa^2 } +f^2 Q^2 +(\nabla \times {\bf Q})^2 \right\rangle.
   \end{equation}
In the presence of vortices ${\bf Q}({\bf r})$ has to be chosen
such that $\nabla \times {\bf Q}$ has the appropriate singularities
along the vortex cores, see e.g.\ Eq.(B4) in App.~B.

  In this paper I consider the ideal periodic FLL in a homogeneous
(pin-free) large superconductor in a uniform magnetic field $H$
along $z$. In this 2D situation one has
$f = f(x,y)$, ${\bf Q} = {\bf Q}(x,y)$, and
${\bf B} = {\bf \hat z} B(x,y)$. Within GL theory in reduced
units the properties of this ideal FLL depend only on two
parameters: the GL parameter $\kappa$ and the average induction
$B = \langle B(x,y) \rangle$.
   The equilibrium magnetic field $H$, and the magnetization
$M = B/\mu_0 - H$, are obtained either from the definition
$H = \partial F / \partial B$ or, more elegantly, from the
virial theorem discovered by Doria, Gubernatis, and
Rainer, \cite{25} which in reduced units reads
   \begin{equation}   
   H ={ \langle f^2 - f^4 + 2\,B(x,y)^2 \rangle \over 2\,B}\,.
   \end{equation}

   Some of the properties of the FLL, and all properties of the
isolated flux line, may be calculated in an elegant way by the
circular cell approximation \cite{7,13,14} as described in App.~A.
In the circular cell method the hexagonal Wigner-Seitz
cell around each flux line is replaced by a circle with radius
$R$ and same area $\pi R^2 = \Phi_0 / B$ if each flux line
carries one quantum of flux
$\Phi_0 = h / 2e = 2.07 \cdot 10^{-^5}$ Tm$^2$. In reduced units
one has $\Phi_0 = 2\pi / \kappa$ and
$R/\lambda = R = (2/b\kappa^2)^{1/2}$ with $b = B/B_{c2}$.
The boundary conditions on the CCM circle $r=R$ are
$df/dr = dB/dr =0$. I find that the free energy of the triangular
FLL, $F_{tr}$, and its magnetization, $M_{tr}$, are reproduced
by the CCM with high accuracy in the entire ranges of $\kappa$
and $B$, $1 / \sqrt2 \le \kappa < \infty$ and $0 \le b < 1$.
In particular, the CCM not only yields $H_{c1}$ (in the limit
$R\to \infty$) but it also reproduces the exact upper critical
field $H_{c2}(\kappa)$, and in the special case
$\kappa = 1 / \sqrt2$ even the exact result
$H(B) = {\rm const} = H_c = H_{c1} = H_{c2}$. These somewhat
surprising features of this approximation are related to the
facts that $H_{c2}$, and in the case $\kappa = 1/\sqrt2$ even
the entire curve $H(B)$, are {\it independent} of the detailed
arrangement of the flux lines, i.e., they are the same for
triangular and square or honey-comb FLLs and for any other
arrangement of single or multiple quanta flux lines.
Another surprising finding is that the virial theorem, Eq.~(3),
works perfectly in the CCM.
Figure 1 shows the magnetization curves $M(H)$ and the
equilibrium field $H(B)$ of the superconductor obtained by
the CCM for $\kappa=0.85$, 1, 1.2, 1.5, 2, 3, 5, 7, 10, and 20.

In the limit $b \to 0$ the CCM yields the lower critical field
$H_{c1}$, which with high accuracy is fitted by the formula
   \begin{eqnarray}  
   \mu_0 H_{c1} = {\Phi_0 \over 4\pi \lambda^2}
   [\, \ln \kappa + \alpha(\kappa) \,] \,, \nonumber
      \\ \nonumber
   h_{c1} = {H_{c1} \over H_{c2}} = { \ln \kappa +
   \alpha(\kappa) \over 2\kappa^2 }  \,,  \\
   \alpha(\kappa) = \alpha_\infty  + \exp[ - c_0
   - c_1 \ln\kappa - c_2 (\ln\kappa)^2] \pm \epsilon
   \end{eqnarray}
with $\alpha_\infty = 0.49693$, $c_0 = 0.41477$,
$c_1 = 0.775$, $c_2= 0.1303$, and $\epsilon \le 0.00076$.
This expression yields at $\kappa = 1/\sqrt 2$ the correct
value $h_{c1} = 1$ and for $\kappa \gg 1$
it has the limit $\alpha = 0.49693$.
A simpler expression for $\alpha(\kappa)$, yielding an
$h_{c1}$ with error still less than 1\% and with the correct
limits at $\kappa=1 / \sqrt2$ and $\kappa \gg 1$, is
  \begin{eqnarray}  
                   \nonumber ~~~~~~~~~~~~~~~\,~~~
  \alpha(\kappa) =0.5 +{1+\ln 2 \over 2\kappa -\sqrt2 +2}\,.
                             ~~~~~~~~~~~~~~ (4{\rm a})
  \end{eqnarray}

    The CCM in principle cannot yield properties related to
the different symmetries of the FLL, or to its shear modulus,
and it cannot give the form factors (Fourier coefficients) of
the magnetic field $B(x,y)$ that may be measured by neutron
scattering. These subtle properties can be computed by the
2D method presented in Ref.~\onlinecite{17} and
in App.~B. This effective numerical method expresses the smooth
functions $f(x,y)^2$ and $B(x,y)$ as 2D Fourier series and
determines the Fourier coefficients by iteration.

   Figure 2 (top) shows the difference of the free energy
densities of the triangular ($F_{tr}$) and square ($F_{sq}$) FLLs.
This difference is proportional to the shear modulus $c_{66}$ of
the triangular FLL (the shear modulus of the unstable square FLL
is negative within GL theory) by the relation \cite{17}
   \begin{equation}   
   c_{66} = (3\pi^2/2)( F_{sq} - F_{tr} ) \,.
   \end{equation}
Note that this difference is very small,
$0 < (F_{sq}-F_{tr})/(\mu_0 H_c^2) < 0.0018$.
Even smaller (by ten times) is the difference between the free
energy densities of the CCM ($F_{cc}$) and of the triangular FLL
plotted in Fig.~2 (bottom). One has
$0 < (F_{cc}-F_{tr})/(\mu_0 H_c^2) < 0.00020$. This result shows
that the CCM is an excellent approximation for global properties
of the FLL. Both differences are largest for large $\kappa$ and
have a maximum near $b \approx 0.3$. The finding $F_{sq}>F_{tr}$
means that the triangular FLL is stable for all
$\kappa > 1/\sqrt 2$. Note that for $\kappa = 1/\sqrt 2$ one has
exactly $F_{sq} = F_{cc} = F_{tr} = 0$ for all $b$.

  Figure 3 (top) shows the difference between the magnetizations
$M_{sq}$ of the square FLL and $M_{tr}$ of the triangular
FLL. Again, this difference is small,
$0.0008 < -(M_{sq}-M_{tr})/ H_{c2}) \le 0.00014$ and the relative
difference has the limits
$-0.018 < (M_{sq}-M_{tr})/ M_{tr} \le 0.0095$.
Figure 3 (bottom) shows the difference between the magnetization
$M_{cc}$ obtained by the CCM (see Fig.~1) and the exact value
$M_{tr}$ of the triangular lattice. Like with the free energy,
this difference is again smaller by a factor of ten than the
difference between two lattice symmetries,
$0.00016 < -(M_{cc}-M_{tr})/ H_{c2} \le 0.00008$ and
$-0.0011 < (M_{cc}-M_{tr})/ M_{tr} \le 0.0017$.
The differences vanish exactly at $\kappa = 1/\sqrt 2$, and
also at $\kappa \to \infty$, since there $m = M/H_{c2} \to 0$.
The {\it relative} differences (insets in Fig.~3) are maximum
at $\kappa \gg 1$.

The smallness of these differences explains why in Fig.~1 the
magnetization curves for all three cases $M_{tr}$, $M_{sq}$,
and $M_{cc}$ coincide within line thickness.
\begin{figure}  
\includegraphics[scale=.515]{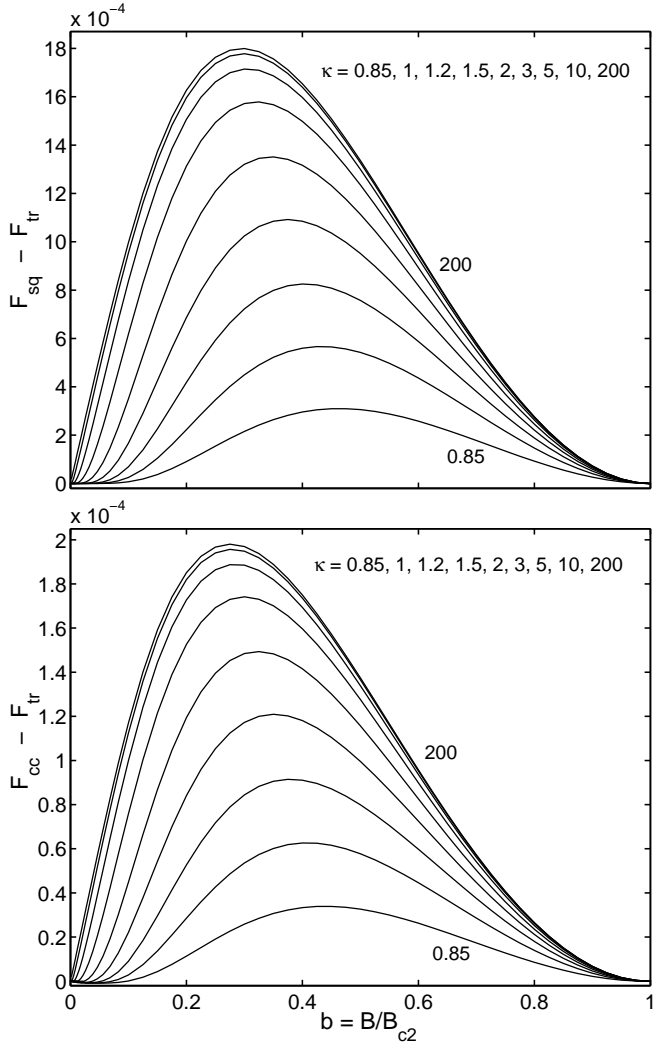}
\caption{\label{fig2} Top: The difference of the free energy
densities of the triangular ($F_{tr}$) and square ($F_{sq}$)
FLLs in units $\mu_0 H_c^2$, plotted versus the reduced
induction $b=B/B_{c2}$ for $\kappa=0.85$, 1, 1.2, 1.5, 2, 3,
5, 10, and 200. This difference equals $(2/3\pi^2)=0.068$
times the shear modulus $c_{66}$ of the triangular FLL.
Bottom: The very small difference between the free energy
densities of the circular cell method ($F_{cc}$) and of the
triangular FLL. Note that the top and bottom
plots look similar, but the scales of the ordinate differ
by a factor of about ten.
 }
\end{figure}
\begin{figure}  
\includegraphics[scale=.505]{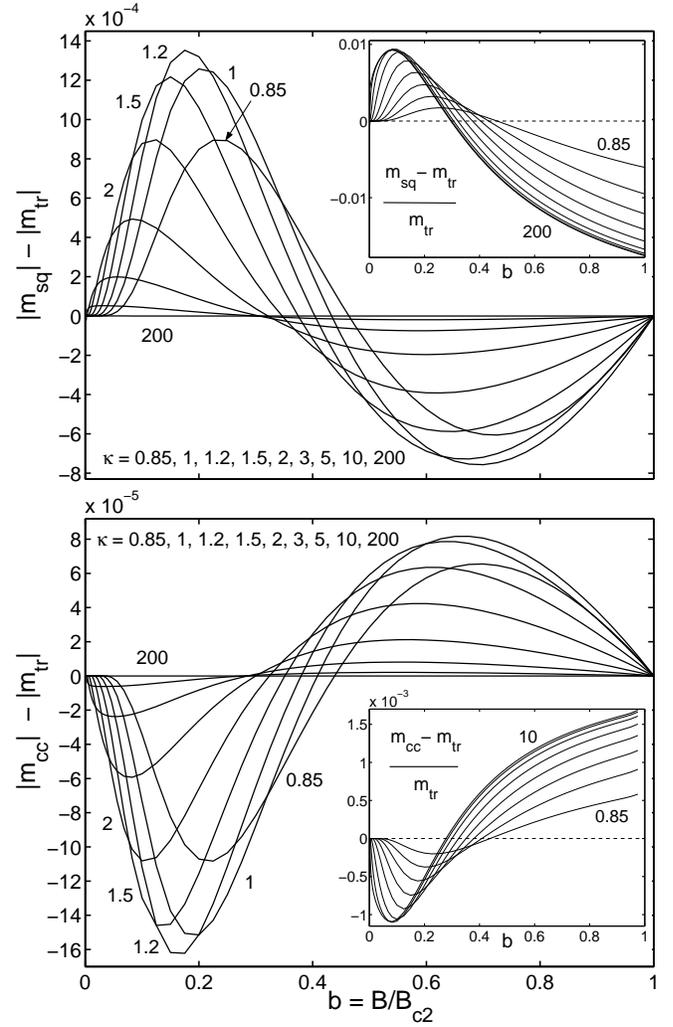}
\caption{\label{fig3} Top: The difference between the
magnetizations $M_{sq}$ of the square FLL and $M_{tr}$ of the
triangular FLL in units $H_{c2}$, plotted versus the reduced
induction $b=B/B_{c2}$ for $\kappa=0.85$, 1, 1.2, 1.5, 2, 3,
5, 10, and 200.  The inset shows the relative difference.
Bottom: The difference between the magnetization $M_{cc}$
obtained by the CCM (see Fig.~1) and the exact value $M_{tr}$ of
the triangular lattice. The inset shows the relative difference.
 }
\end{figure}
\begin{figure}  
\includegraphics[scale=.48]{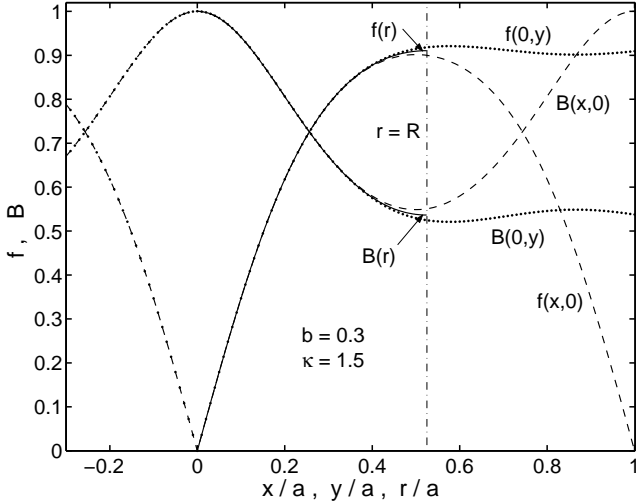}
\caption{\label{fig4} Comparison
of the GL functions $f$ and magnetic inductions $B$
calculated for the triangular FLL and from the circular cell
approximation for the example $b=0.3$, $\kappa = 1.5$.
Shown are the cross sections $f(x,0)$, $B(x,0)$ along the
nearest neighbor direction $x$, and $f(0,y)$, $B(0,y)$ along
the perpendicular direction $y$, and $f(r)$, $B(r)$ from the
CCM. All $B$ are in units $B(0,0)$ of the triangular FLL.
Small deviations can be seen only close to the cell
boundary $r=R$, $R/a = 3^{1/4} (2\pi)^{-1/2} = 0.525$. At
lower $b$ the deviations are even smaller.
 }
\end{figure}
\begin{figure}  
\includegraphics[scale=.50]{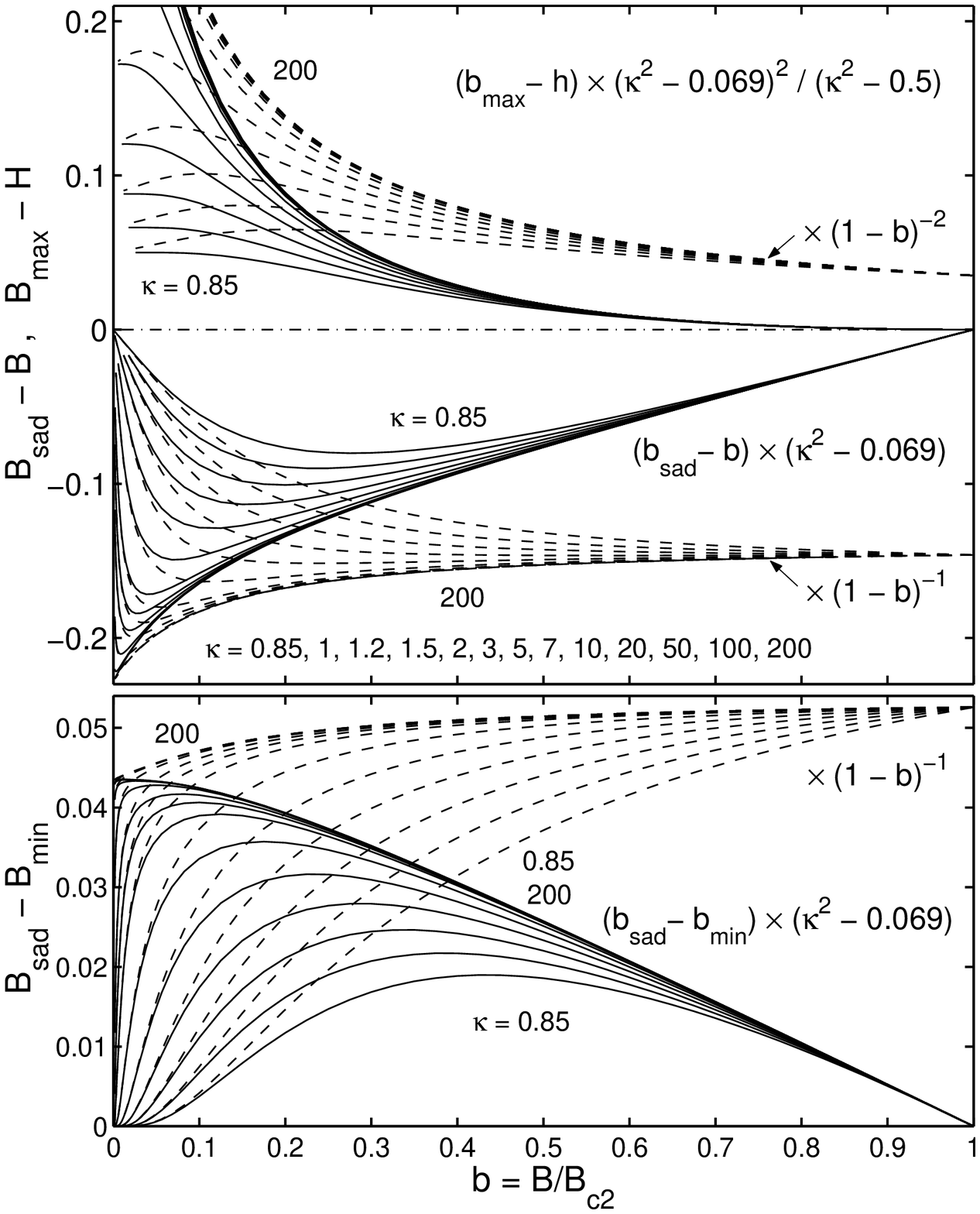}
\caption{\label{fig5}
  Maximum field $B_{max} =b_{max} B_{c2}$ minus applied field
$H$, saddle-point field $B_{sad} =b_{sad} B_{c2}$ minus
induction $B$, and $B_{sad}$ minus minimum field
$B_{min} = b_{min} B_{c2}$, for the triangular FLL,
plotted versus $b=B/B_{c2}$ for
$\kappa = 0.85$, 1, 1.2, 1.5, 2, 3, 5, 7, 10, 20, 50, 100, 200.
The solid lines show these small differences in units
$B_{c2}$, multiplied by appropriate functions of $\kappa$ to
obtain collapse of the curves near $b=1$.
The dashed lines show the same functions multiplied by factors
$(1-b)^{-2}$ and $(1-b)^{-1}$ such that they tend to a finite
constant value near $b=1$, cf.\ Eqs.~(6) - (8).
 }
\end{figure}

   Figure 4 shows an example ($b=0.3$, $\kappa = 1.5$)
comparing the spatial functions $f$ and $B$ of the triangular
FLL and from the CCM. Shown are the cross sections $f(x,0)$
along the nearest neighbor direction $x$ and $f(0,y)$
perpendicular to this, and $f(r)$ from the CCM [\,$a$ is the
vortex spacing, $a^2/\lambda^2 =4\pi/(\sqrt3 b \kappa^2)$\,].
It is seen that
$f(x,0)$ and $f(r)$, and also $B(x,0)$ and $B(r)$, coincide
closely; at lower $b< 0.3$ the difference is smaller than the
line thickness. The solutions for the square FLL deviate more
from the circular cell solutions.

  The maximum, minimum, and saddle-point fields of the
triangular FLL, $B_{max}=B(0,0)$, $B_{min}=B(0, a/\sqrt3)$, and
$B_{sad}=B(a/2,0)$, depend on $b$ and $\kappa$.
$B_{max}$ is only slightly above the equilibrium field $H$, and
$B_{sad}$ and $B_{min}$  are close to each other and lie
somewhat below the average field $B$. $B_{max}$ and $B_{min}$
are shown in Fig.~3 of Ref.\ \onlinecite{17}
as functions of $b$ for several $\kappa= 0.707 \dots 5$.
In Fig.~5 the small differences $B_{max}-H$, $B_{sad}-B$,
and $B_{sad}-B_{min}$ are plotted versus $b$, in units $B_{c2}$
and multiplied by a function of $\kappa$ such that the curves
for all $\kappa \ge 1/ \sqrt2$ collapse at $b \to 1$.
One finds for all $\kappa$ near $b=1$:
   \begin{eqnarray}   
   {B_{max}-H \over B_{c2}} \approx 0.0351\,
    {\kappa^2-0.5 \over (\kappa^2 -0.069)^2} (1-b)^2 \,, \\
   {B_{sad}-B \over B_{c2}} \approx -0.146 \, {1-b \over
   \kappa^2 -0.069}  \,, \\
   {B_{sad}-B_{min} \over B_{c2}} \approx 0.0526\, {(1-b)
   \over \kappa^2 -0.069}  \,.
   \end{eqnarray}
The factor 0.069 in Eqs.~(6)--(8) is $0.5-0.5/\beta_A =0.0688$
where $\beta_A = 1.1596$ is the Abrikosov parameter of the
triangular FLL. Plots of $B_{cc}(R) - B_{min}$ where
$B_{cc}(R)$ is the field value at the boundary of the circular
cell in the CCM, look similar to the plots of $B_{sad}-B_{min}$
in Fig.~5 (lower panel), since the value $B_{cc}(R)$ lies
approximately in the middle between $B_{min}$ and $B_{sad}$,
see Fig.~4. Since for $\kappa \gg 1$ and $b\ll 1/\kappa^2$ the
field in the vortex center equals $B_{max} = 2 H_{c1}$,
one has $B_{max} - H \to H_{c1}$, and thus the function
plotted in Fig.~5 (upper panel) for $b\to 0$ tends to the
limit $(b_{max}-h) \times \kappa^2 \to h_{c1} \kappa^2 \approx
{1\over 2}(\ln \kappa +0.50)$, cf.\ Eq.~(4).
\begin{figure}  
\includegraphics[scale=.475]{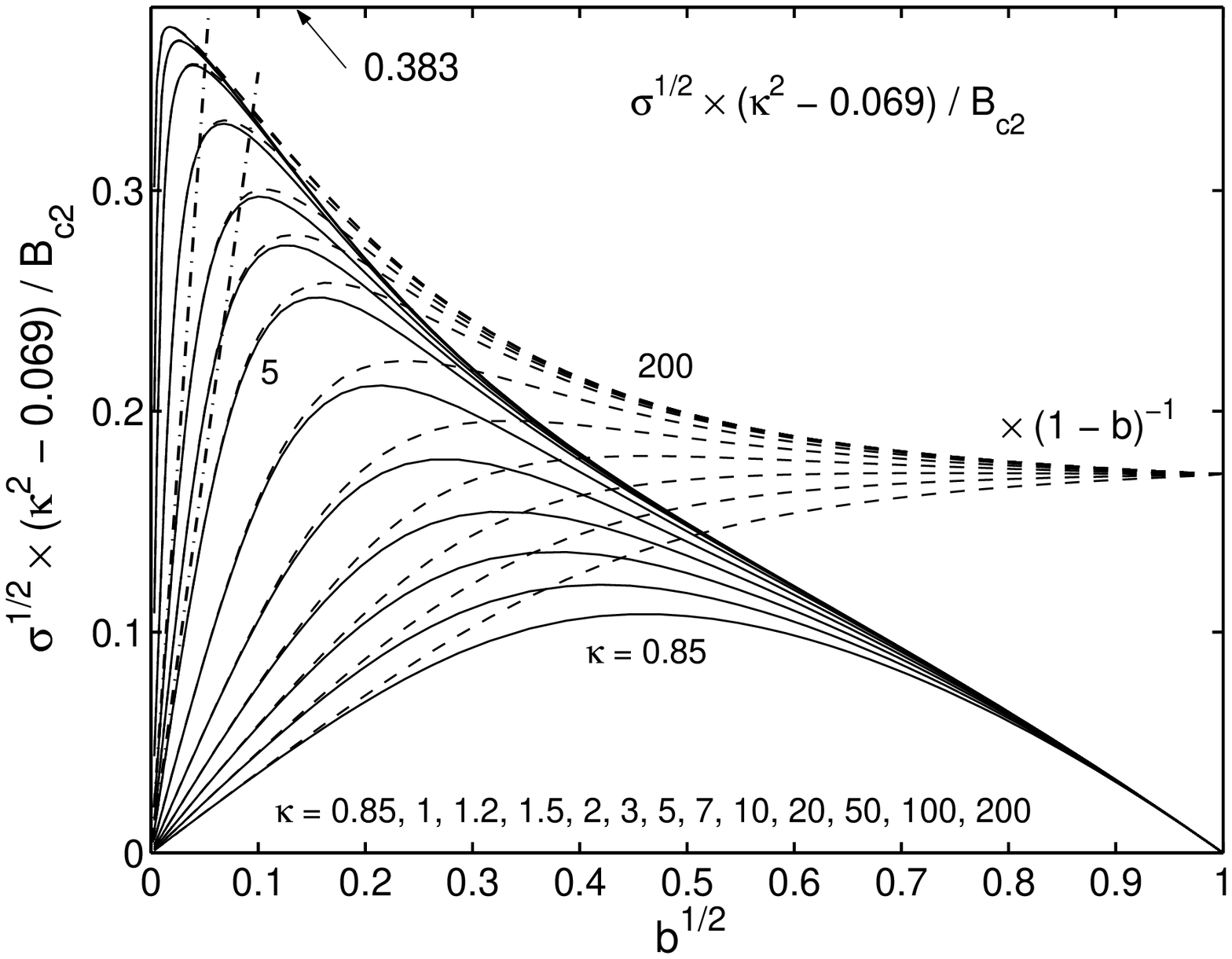}
\caption{\label{fig6}  The magnetic field variance
$\sigma = \langle [B(x,y) - B]^2 \rangle$ of the triangular FLL
for $\kappa=0.85$, 1, 1.2, 1.5, 2, 3, 5, 7, 10, 20, 50, 100, 200
plotted in units of $B_{c2}$ as
$\sqrt{\sigma}\cdot(\kappa^2-0.069)/B_{c2}$ (solid lines) such that
the curves for all $\kappa$ collapse near $b=1$, cf.\ Eq.~(10).
The dashed lines show the same functions divided by $(1-b)$
such that they tend to a finite constant 0.172 at $b=1$.
All curves are plotted versus $\sqrt b = \sqrt{B/B_{c2}}$ to
stretch them at small $b$ values and show that they go to
zero linearly. The limits, Eq.~(12), for $\kappa=5$ and
$\kappa =10$ are depicted as dash-dotted straight lines.
The upper frame 0.383 shows the approximation (11).
 }
\end{figure}
\begin{figure}  
\includegraphics[scale=.535]{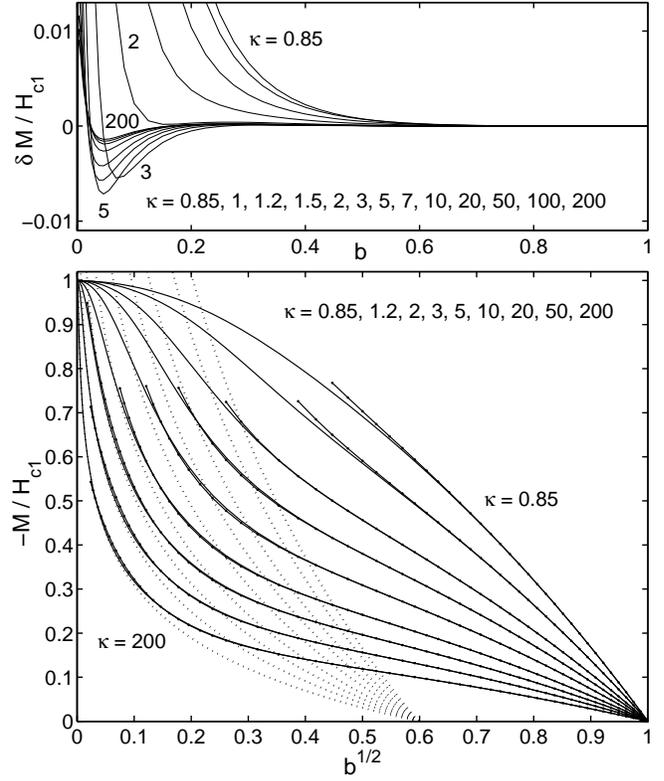}
\caption{\label{fig7} Lower panel: The exact magnetization
$M$ of the triangular FLL (solid lines) and the fit,
Eq.~(15), (solid lines with dots) plotted for many $\kappa$
values versus $\sqrt b = \sqrt{B/B_{c2}}$ to stretch the
low field region. Shown is $-M$ normalized to its maximum
value $H_{c1}$ occurring at $b=0$. The fit (15) is good for
all $\kappa$ and not too small $b > 1/(4\kappa^2) +0.0005$.
Upper panel: The deviation $\delta M$ of the fit from the
exact $M$ is very small when $b>0.5$.
The dotted lines in the lower panel show the old London
approximation, Eq.~(18).
 }
\end{figure}
\begin{figure}  
\includegraphics[scale=.485]{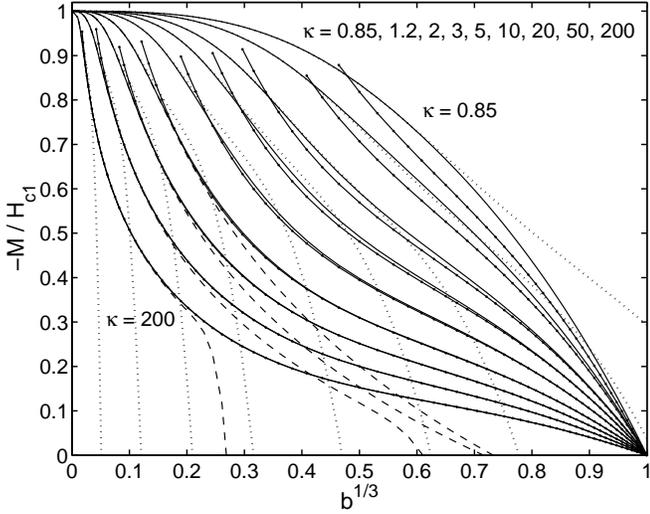}
\caption{\label{fig8} Exact magnetization of the triangular
FLL (solid lines) and the logarithmic fit, Eq.~(19),
(solid lines with dots) plotted versus $b^{1/3}$ to stretch
the region at small $b=B/B_{c2}$.
The dotted lines show the London nearest-neighbor
approximation, Eq.~(20).
The dashed lines show the London expression, Eq.~(C6), with
the sum taken over all shells up to $\nu_{max} =100$ vortex
spacings. Both London approximations are good fits at very
low $b$ and all $\kappa$.
 }
\end{figure}

   The variance of the magnetic field is
   \begin{eqnarray}   
   \sigma =\langle[ B(x,y)-B ]^2 \rangle
          =\langle B(x,y)^2\! -B^2  \rangle
          =\!\sum_{{\bf K} \ne 0}\!\! B_{\bf K}^2 \,,
   \end{eqnarray}
where $B_{\bf K}$ are the Fourier coefficients of
$B(x,y)=\sum_{\bf K} B_{\bf K} \cos {\bf Kr}$ and ${\bf K}$ the
vectors of the reciprocal lattice of the FLL (App.~B). Near
$b=1$ the Abrikosov solution of the linearized GL
theory \cite{11,26} yields for all $\kappa$ values \cite{27}
   \begin{eqnarray}   
   \sigma = 7.52\cdot10^{-4} \, {\Phi_0^2 \over \lambda^4} \,
   {\kappa^4 ~(1-b)^2 \over (\kappa^2 -0.069)^2} \,, \nonumber \\
   S \equiv {\sqrt\sigma \over B_{c2}} =
   0.172\,{1-b \over \kappa^2 - 0.069} \,.
   \end{eqnarray}
The functions $S$ and $S/(1-b)$ are plotted in Fig.~6 versus
$\sqrt b$ for various $\kappa$. It can be seen that Eq.~(10)
is a rather good approximation for the large range
$0.25 < b < 1$. At smaller $b$ the variance $\sigma(b)$ has a
maximum and then goes to zero again at $b=0$.

  For small inductions $b\ll 1$ and large $\kappa$ one can use
the London approximation $B_{\bf K} = B/(1 +K^2 \lambda^2)$.
For the appropriate cutoff at large magnitudes
$K \sim \xi^{-1} =\kappa/\lambda$ see Ref.~\onlinecite{28,29}
and below. In the range $0.13/\kappa^2 \ll b \ll 1$ the unity
in the denominator of $B_{\bf K}$ may be disregarded since
$K^2\lambda^2 \ge (4\pi/\sqrt3) b \kappa^2 = 7.255 b \kappa^2$.
Thus $B$ drops
out and $\sigma$ becomes independent of $b$: \cite{27}
   \begin{eqnarray}   
   \sigma = 0.00371 {\Phi_0^2 \over \lambda^4},~~~
   S ={0.383 \over \kappa^2} \,.
   \end{eqnarray}
This often used formula corresponds to the upper axis
in Fig.~6. One can see that this approximation is good
only for very large $\kappa \ge 70$ and only in the range
of $b$ near the maximum of $\sigma$. At very small
$b \ll 0.13/\kappa^2$ both $\sigma(b)$ and $S(\sqrt b)$
drop linearly to zero when $b\to 0$. In this range the sum
in Eq.~(9) can be evaluated as an integral, yielding
   \begin{eqnarray}   
   \sigma ={b\kappa^2 \over 8\pi^2}\,{\Phi_0^2 \over\lambda^4},
   ~~~ S ={\sqrt{b/2} \over \kappa} \,.
   \end{eqnarray}
This approximation is good for $\kappa \ge 5$ and very small
$b$ ($b< 0.01/\kappa^2$ for $\kappa =5$; $b <0.04/\kappa^2$
for $\kappa \ge 10$), see the two straight lines in Fig.~6.
For $\kappa \ge 5$ a good approximation, with less than 5\% error
from $b=1$ down to the value $b \approx 0.25/\kappa^{1.3}$
where the maximum of $\sigma$ occurs, is
   \begin{equation}   
   S \equiv {\sqrt\sigma \over B_{c2}} \approx 0.172\,
   {1-b \over \kappa^2} [\,1 +1.21(1-\sqrt{b})^3 \,] \,.
   \end{equation}
This approximation is much better that the interpolation,
Eq.~(3) of Ref.~\onlinecite{27}.

\section{Magnetization curves}    

  This section presents analytic expressions which approximate
the computed magnetization $m = M/H_{c2} =b-h$ (Fig.~1) as a
function of the induction $b=B/B_{c2}$ or of the thermodynamic
field $h=H/H_{c2}$. We distinguish approximations working at
high or low inductions.

\subsection{Approximation for high inductions}  

  The linearized GL theory yields for $1-b \ll 1$  Abrikosov's
$B_{c2}$ solution \cite{1,11}
   \begin{equation}  
   m \approx m_A = - {1-b \over (2\kappa^2 -1) \beta_A +1 } \,,
   \end{equation}
where $\beta_A =\langle \omega_A^2 \rangle / \langle
\omega_A \rangle^2 = 1 + \sum_{m,n} \exp[\, K_{mn}^2
 S/(4\pi) \,]$ (Ref.\ \onlinecite{11,30} and App.~B) is
the Abrikosov parameter, $\beta_A = 1.1595953$
($\beta_A=1.1803406$) for the triangular (square) FLL.
The linear magnetization $m_A(b,\kappa)$ is a good
approximation in the range $0.5 \le b \le 1$, see
Fig.~1. This suggests the following fit to the exact $m$:
   \begin{eqnarray}  
   m(b,\kappa) = m_A - (1-b)^2 \exp[\,f_1(b)\,]\, g_1(\kappa)
    +\epsilon_1 , \nonumber \\
   f_1(b) = 2.50u^2-8.08u+0.39, ~ u=(1-b)^{0.41} ,
      \nonumber \\
   g_1(\kappa) = ( 1.133 +1.926 / \kappa^{2.25} )
     ( 2\kappa^2  -1)/( 2\kappa^4 ) \,.\,
   \end{eqnarray}
with relative error $| \epsilon_1 / m| < 0.0013$ for $b>0.5$
for the triangular FLL.
Formula (15) is a good approximation with relative
error $< 1 \%$ for all $\kappa$ in the large range of fields
$ 1/(4 \kappa^2) + 5 \cdot 10^{-4} \le b \le 1$, see Fig.~7.

The same expression (14) fits also the $m(b,\kappa)$ of
the square FLL, with somewhat larger error if the same
functions $f_1(b)$,  $g_1(\kappa)$ are used rather than
the optimally fitted ones. For the difference
$m_{\rm tr} - m_{\rm sq}$ see Fig.\ 3.

\subsection{Approximation for ``intermediate fields''} 

  For completeness I mention here also the London
approximation \cite{3} which was supposed to be good in
the ``intermediate field range'' $H_{c1} \ll H \ll H_{c2}$
that exists only in superconductors with extremely
large $\kappa$. Within London theory the induction is
(see App.~B)
  \begin{eqnarray}  
  B(x,y) =B \sum_{\bf K} {\cos{\bf Kr}
  \over 1+K^2 \lambda^2} \,,
  \end{eqnarray}
where the sum goes over all $K$-vectors with length from
$K=0$ to some cutoff $K \approx \xi^{-1}$. Inserting this
into the London free energy density
$[B({\bf r})^2 + \lambda^2 (\nabla \times {\bf B})^2] /(2\mu_0)$
and averaging over the superconductor one gets
  \begin{eqnarray}  
  F = \sum_{\bf K} { (B^2/2\mu_0) \over 1 +K^2 \lambda^2 }
    \approx {B^2 \! \over 2\mu_0} +{B \Phi_0 \over 2\mu_0}
    \int {d^2 k \over 4 \pi^2} {1 \over k^2 \lambda^2} \,.
  \end{eqnarray}
The integral from
 $k_{\rm min}^2 \approx (K_{10}/2)^2 \approx \pi^2 B/\Phi_0$
to $k_{\rm max}^2 \approx \xi^{-2} =2\pi B_{c2}/\Phi_0$ equals
$(4\pi \lambda^2)^{-1} \ln(\gamma'/b)$ where $\gamma'$ is
some constant and $b=B/B_{c2}$ as above. This yields
  \begin{eqnarray}  
  -M &=&H -{B\over \mu_0} ={ \partial F \over \partial B} -
  {B \over \mu_0} ={\Phi_0 \over 8\pi\lambda^2 \mu_0}
  \ln{\gamma\over b} \,, \nonumber \\
  -m &=& {-M \over B_{c2}} = {1 \over 4\kappa^2 }
  \ln{0.358 \over b}
  \end{eqnarray}
with constant $\gamma = \gamma'/e = 0.3575\dots$ obtained
by our fit to the numerical $m(b)$ at $\kappa=200$.
This old London approximation is shown in Fig.~7 as
dotted lines. One sees that this fit works only at large
$\kappa \ge 20$ in the relatively small interval
$1/(2 \kappa^2) \le b \le 0.01$, i.e.\ at very low $b$
(but not too low $b$).
It gives $m=0$ at $b=\gamma$ for all $\kappa$.
This fit is slightly improved by replacing
$\ln(\gamma/b)$ by $\ln(1-\gamma +\gamma/b)$, which
gives the correct $m=0$ at $b=1$.

  A much better fit in the spirit of this logarithmic
approximation is (see Fig.~8)
  \begin{eqnarray}  
 -m &=& {1 \over 4\kappa^2 } \ln\left[ 1+ {1-b \over b}
    f_2(b) \right] \,,  \nonumber \\
   f_2(b) &=& 0.357+2.890\,b-1.581\,b^2 \,.~~
  \end{eqnarray}
This fit is good for $\kappa \ge 3$  (error $ < 3\%$) and
$\kappa \ge 5$ (error $ < 1\%$ ) in the large ranges
$ (\ln\kappa +1)/(10\kappa^2) \le b \le 1$ for
$\kappa = 3 \dots 200$.
These intervals of validity may also be expressed as
$-M/H_{c1} = -m/h_{c1} \le 0.8$ (0.85) for $\kappa \le 20$
($\kappa \ge 50$).

\subsection{Approximation for low inductions} 

   All the above approximations do not describe the correct
vertical slope of $M(H)$ at $H=H_{c1}$, or zero slope of
$H(B)$ and unity slope of
$M(B) = B-H$ at $B=0$. This is achieved by the London
approximation of pairwise interacting vortices described
in App.~C. For very small $b \ll 1$ one may account only
for the nearest neighbor shell of six vortices in the
triangular FLL of spacing $a =c\lambda$. With $h(b)$,
Eq.~(C8), this yields for $-m(b) = h(b) -b$:
  \begin{eqnarray}   
  -m \approx h_{c1} -b + {3 \sqrt{\pi c} \over 2 \kappa^2}
  e^{-c}  \left[1+{19\over8c} -{47\over128c^2} \right] ,
   \nonumber \\
  c = {a \over \lambda} = \left( {4\pi / \sqrt3 \over
    b \kappa^2 } \right)^{1/2} .
  \end{eqnarray}
Formula (20) correctly describes the steep diverging slope of
$m(h) \to \infty$ or slopes $m(b)' \to 1$ and $h(b)' \to 0$
as $b \to 0$ and is valid for $0 \le b \le 2.5/\kappa^2$ for
$\kappa \ge 7$, Accidentally it also fits well $m(b)$
for $\kappa \le 2$ and $b \le 0.2$, see the dotted lines
in Fig.~8. A smoother fit is
obtained by the exact London expression (C6) if
one or three neighbor shells are included in the sum.
But taking more terms in the sum improves the fit only
at large $\kappa$.
Accounting for neighbors up to $\nu =100$ lattice spacings
apart (about 5000 terms) one gets good approximation
to $m$ and $h$ for $0 \le b \le 0.01$ (0.02, 0.05)
if $\kappa \ge 20$ ($\kappa =7$, $\kappa = 2$), see Fig.8.
In the limit $\nu \to \infty$ the infinite sum (C6)
reproduced Eq.~(18), i.e, the dashed curves in Fig.~8
for $\kappa = 50,\, 200$ then will straighten and cut
the axis $M=0$ at $b = \gamma = 0.358$
($b^{1/2} = 0.60$ in Fig.~7, $b^{1/3} = 0.71$ in Fig.~8).

\subsection{General interpolation}  

  All the approximations for $m(b)$ and $h(b)$ known
so far, including the above formulas, fit either
the low or high field regions. The formulas (15) and (20)
[or better, Eq.~(C6) with the sum taken over three shells]
have a small overlap for all $\kappa$ and thus, together,
they fit the entire range $0\le b \le 1$ [though the
good fit of the low-$\kappa$ data by the London expression
(20) or (C6) is accidental].

   For practical purposes one may construct interpolation
formulas that approximate the numerically obtained
magnetization in the entire range $0<b<1$. They should
satisfy the five conditions
  \begin{eqnarray}   
  h(0)=h_{c1},~ h'(0)=0,~ h(1)=1,~ h'(1)=1-p\,, \nonumber \\
  h''(1)=0,~~~p=m'(1)=[\,(2\kappa^2-1) \beta_A +1\,]^{-1}\,,~
  \end{eqnarray}
with $h_{c1}(\kappa)$ from Eq.~(4). A simple expression
that satisfies all these conditions is
  \begin{eqnarray}   
  - m(b,\kappa) = h-b = p\,(1-b) + (h_{c1}-p)(1-b)^\eta
  \end{eqnarray}
with $\eta(\kappa) = (1-p)/(h_{c1}-p)$.
Formula (22) approximates the exact $-m(b)$ well for
$\kappa \le 2$ with relative deviation $|\epsilon| < 3\%$,
for $\kappa = 3$ with $-2\% < \epsilon < 6\%$,
and for $\kappa = 5$ with $ -1\% < \epsilon < 16\%$,
see the dotted lines in Fig.~1, top.

For large $\kappa$, general interpolation formula are more
difficult to construct because of the nonanalytic
limiting expression Eq.~(20). One may, however,
combine the $m_{\rm low}$ from Eq.~(20) with the
$m_{\rm high}$ from Eq.~(19) using a smooth transition
at $b \approx (2\kappa^2)^{-1}$, e.g., with weights
$1-w$ and
$w={1\over2}+{1\over2} \tanh[\,2.5(2b\kappa^2 -1)\,]$,
or slightly better,
$w={1\over2}+{1\over2} {\rm \,erf\,} [\,2(2b\kappa^2-1)\,]$,
yielding
  \begin{eqnarray}   
   m(b,\kappa) = (1-w)\, m_{\rm low}
    + w \, m_{\rm high} \,.
  \end{eqnarray}
This interpolation between expressions (19) and (20) works
well for $0 < b < 1$ with relative error $|\epsilon| < 2\%$
for $\kappa \ge 5$ and $-3.5\% < \epsilon < 2\%$ for
$\kappa \ge 1$, see the dotted lines in Fig.~1 (bottom).
Thus, $m$ in the entire ranges of $b$ and $\kappa$ may
be approximated by Eq.~(22) or Eq.~(23).

\section{Conclusion}  

  The properties of the ideal periodic flux-line lattice
in superconductors are calculated from Ginzburg-Landau
theory for the entire ranges of GL parameters
$1 / \sqrt2 \le \kappa < \infty$ and inductions
$0 \le b=B/B_{c2} < 1$. The differences between the
free energies and magnetizations of the triangular and
square vortex lattices and the values obtained by the
circular cell approximation are investigated in detail.
Approximate analytical expressions are given for the
variance $\sigma(b, \kappa)$ of the periodic induction
and for the magnetization $m(b, \kappa)$. These limiting
and interpolation formulas should replace previous
approximate expressions that have rather limited validity.

The numerical methods presented in the appendices, in
principle may be applied also to theories going beyond
the isotropic GL theory considered here.

\appendix 
\section{Isolated vortex and circular cell method}

   The calculation of the isolated flux line and of the
FLL within the circular cell method, is
a cylindrically symmetric problem. The free energy depends
on the magnitude of the GL function $f(r)$ and on the magnetic
induction $B(r)$ (along $z$) related to the vector potential
$A(r)$ and supervelocity $Q(r)$ (along $\varphi$) by
\begin{equation} \label{A1}
   B = {(A r)' \over r} = {(Q r)' \over r} ,
   ~~ Q=A -{1\over \kappa r} \,.
\end{equation}
In reduced units $\sqrt2 H_c = \mu_0 H_c^2 = \lambda = 1$,
the free energy of a flux line or of the circular cell with
radius $R$  ($\pi R^2 = \Phi_0/B$) averaged over this cell
and referred to the Meissner state ($f=1$, $B=0$) reads
\begin{equation} \label{A2}
  F_{cc} \!=\!\int_0^R \!\left[{(1\!-\!f^2)^2 \over2}
   +{(f')^2 \over \kappa^2} + \! f^2Q^2 \!
   + \!B^2 \right] {2\pi r\,dr \over \pi R^2}
\end{equation}
with $f' = df/dr$. Minimizing the functional (A2) with
respect to $f(r)$ and $Q(r)$ we obtain the two GL equations,
which may be written in the form
  \begin{eqnarray}   
  -f'' +\kappa^2 f = \kappa^2 (2f -f^3 -Q^2 f) +f'/r \,,\\
  B' = f^2 Q = j \,,\,
  \end{eqnarray}
where $j= B'$ is the current density. In Eq.~(A3) a term $\kappa^2 f$
was added on both sides to improve the convergence of the
iteration below. The boundary conditions are
\begin{equation} 
  f(0) = f'(R)= j(0) = j'(R) =0.
\end{equation}
An appropriate ansatz in terms of Fourier series is
  \begin{eqnarray} 
 f(r)= \sum_{m=1}^M f_G \sin Gr, ~~~G={\pi(2m-1) \over 2R} \,,~\\
 A(r)= \sum_{n=1}^N a_K \sin Kr + {r\over 2} B, ~~\,~
         K = {\pi n \over R} \,,~\\
 B(r)= \sum_{n=1}^N a_K {\sin Kr + Kr \cos Kr \over r} ~+B \,,~\\
 Q(r)= \sum_{n=1}^N a_K \sin Kr~~-~{1-r^2/R^2 \over\kappa r} \,,~\\
 j(r)=\!\sum_{n=1}^N \!a_K {Kr \cos Kr
       - (1+K^2r^2)\sin Kr \over r^2} \,.~
  \end{eqnarray}
For the equidistant grid $r_i = (i-{1\over 2})R/N_r$,
$i=1, 2, \dots N_r$, one has the orthogonality relation
  \begin{equation}  
  \sum_{i=1}^{N_r} \sin Gr_i \sin G'r_i
   = {1\over 2} \, N_r \, \delta_{G G'}
  \end{equation}
and similar equations for $\sin Kr_i$ and $\cos Kr_i$. The GL
equations (A3) and (A4) thus may be written in the form of
equations for the Fourier coefficients $f_G$ and $a_K$:
  \begin{eqnarray}  
  f_G = {1\over G^2 +\kappa^2}{2\over N_r} \sum_{i=1}^{N_r}
   \sin Gr_i \times  \nonumber \\ {
   [ \kappa^2(2f-f^3-Q^2f)+f'/r_i] } \,, \\
  a_K ={1\over K^2 +1} \Big[ a_K + {2\over N_r} \sum_{i=1}^{N_r}
   \sin Kr_i \times   \nonumber \\
   \Big( \sum_{n'=1}^N  a_{K'} {Kr\cos Kr - \sin Kr \over r^2}
   -f^2 Q \Big) \Big] \,.
  \end{eqnarray}
These two equations may be used to obtain the $f_G$ and $a_K$
by iteration, starting with appropriate initial values.
The iteration becomes more stable and faster if the
value of the previous iteration step is added with a certain
weight $(1-c) < 1$, e.g., c=0.6, according to the algorithm:
  \begin{eqnarray}  
  f_G \leftarrow (1-c)f_G +\,c F_G \,\{ f,Q \} \,, \\
  a_K \leftarrow (1-c)a_K +c A_K \{ f,Q \} \,,
  \end{eqnarray}
with the symbols $F_G \{f,Q\}$ and $A_K \{f,Q\}$ denoting
the right hand sides of Eqs.~(A12) and (A13), respectively.
Rapid convergence is achieved by iterating equations (A14),
(A15) alternately. The equilibrium magnetic field $H$ is
then obtained from Eq.~(3), and the magnetization from
  \begin{eqnarray}  
  M = {2\over B R} \int_0^R \! \Big[{f^4 -f^2 \over 2} +B^2
    -B(r)^2 \,\Big]  r \,dr \,.
  \end{eqnarray}

  At very large $\kappa$ and very small $b$ a large number
$N_r$ of grid points $r_i$ is needed to achieve high accuracy,
$N_r \gg R/\xi = R \kappa =\sqrt{2/b}$. In this case the
accuracy with a limited
number of grid points may be improved by choosing a
nonequidistant grid, e.g., $r_i = u_i^2$ with equidistant
$u_i= (i-{1\over 2}) \sqrt R /N_r$. To use the orthogonality
relations one then has to express $f$, $B$, and $Q$ as
Fourier series in the new variable $u=r^2$ and also write
the two GL equations in terms of the variable $u$,
using, e.g., $f'(r) = f'(u)/2u$ and
$f''(r) = f''(u)/4u^2 - f'(u)/4u^3$. This yields
  \begin{eqnarray} 
 f''(u) = 4 u^2 \kappa^2 (-f+f^3+Q^2 f) +f'/u \,, \\
 B'(u) = 2u f^2 Q \,, \,
  \end{eqnarray}
and the Fourier series
   \begin{eqnarray} 
  f(u)= \sum_{m=1}^M f_G \sin Gu, ~~~G={\pi(2m-1) \over 2R}\,,~\\
  A(u)= \sum_{n=1}^N a_K \sin Ku + {u^2\over 2} B, ~~~
          K = {\pi n \over R} ~\,,~\\
  B(u)= \sum_{n=1}^N a_K { 2\sin Ku +Ku \cos Ku \over
          2 u^2}  +B \,,~\\
  Q(u)= \sum_{n=1}^N a_K \sin Ku~-{1-u^4/R^2 \over\kappa u^2} \,,~\\
  j(u)= \!\sum_{n=1}^N \! a_K {Ku \cos Ku \!-\!
         (4 \!+\! K^2u^2)\sin Ku \over 4u^4} \,.~
  \end{eqnarray}
The equations for the new Fourier coefficients are
  \begin{eqnarray}  
  f_G = {1\over G^2 + 4\kappa^2}\Big[ 4\kappa^2 f_G +{2\over N_r}
   \sum_{i=1}^{N_r}  \sin Gu_i \times  \nonumber \\ { \Big( 4
   u_i^2 \kappa^2 f\, (1-f^2-Q^2) +f'/u_i\Big) } \Big] \,,\,~
                                \\
  a_K ={1\over K^2 +1} \Big[ a_K + {2\over N_r} \sum_{i=1}^{N_r}
   \sin Ku_i \times   \nonumber \\
   \Big( \! \sum_{n'=1}^N \!a_{K'} {Ku_i\cos Ku_i - 4\sin Ku_i
   \over u_i^2} -4 u^2 Q f^2 \Big) \Big] \,.\,~
  \end{eqnarray}
For better convergence a term $-4\kappa f_G$ was added on both sides
of Eq.~(A17) to yield (A24). The corresponding iteration scheme
using (A14), (A15) needs a smaller weight $c$ and more iteration
steps, but for large $\kappa^2/b$ it is faster than the first
scheme since it needs less grid points $N_r$ to reach the same
accuracy.

 \section{Periodic vortex lattice}

   The properties of the ideally periodic FLL within GL theory
may be calculated by minimizing the GL free energy of the
superconductor, Eq.~(2), with respect to appropriate periodic
trial functions, e.g., Fourier series with a large number
of terms. For the smooth function $\omega = f^2({\bf r})$ we
write the ansatz
  \begin{eqnarray}  
  \omega({\bf r}) =
  f^2\!\! &=& \sum_{\bf K} a_{\bf K} (1 -\cos{\bf K r})
  \end{eqnarray}
with  ${\bf r} = (x,y)$, ${\bf K} = (K_x,K_y)$.
In all sums here and below the term ${\bf K} =0$ is
excluded. For vortex positions
${\bf R} = {\bf R}_{mn} = (m x_1 +nx_2, \, ny_2)$
the reciprocal lattice vectors are
${\bf K} = {\bf K}_{mn} = (2\pi/S)(my_2, \,nx_1 +mx_2)$
with $S= x_1 y_2 = \Phi_0/B$ the unit cell area and
$m, n = 0, \pm 1, \pm 2, \,\dots$~.
For the triangular lattice one has $x_2 = x_1 /2$,
$y_2 = x_1 \sqrt 3/2$, and for the square lattice $x_2 = 0$,
$y_2 = x_1$.
For supervelocity ${\bf Q}$ and induction
${\bf B} =\nabla\! \times\! {\bf Q} =B({\bf r}) {\bf \hat z}$
we choose
 \begin{eqnarray}    
  B({\bf r}) &=& B + \sum_{\bf K} b_{\bf K} \,\cos{\bf K r}\,,\\
  {\bf Q(r)} &=& {\bf Q}_A({\bf r}) + \sum_{\bf K} b_{\bf K}
  {{\bf\hat z \times K} \over K^2} \sin{\bf K r} \,.
  \end{eqnarray}
Here  ${\bf Q}_A(x,y)$ is the supervelocity of the
Abrikosov $B_{c2}$ solution, which satisfies
  \begin{eqnarray}    
  \nabla \times {\bf Q}_A = \Big[ B - \Phi_0 \sum_{\bf R}
  \delta_2({\bf r-R}) \Big] {\bf\hat z}\,,
  \end{eqnarray}
where $\delta_2({\bf r}) = \delta(x)\delta(y)$ is the 2D delta
function. This relation shows that ${\bf Q}_A$ is the velocity
field of a lattice of ideal vortex lines but with zero average
rotation. Close to each vortex center one has
${\bf Q}_A({\bf r}) \approx {\bf\hat z \times r'}/(2\kappa r'^2)$
and $\omega({\bf r}) \propto r'^2$ with ${\bf r'= r-R}$.
In principle ${\bf Q}_A({\bf r})$ may be expressed as a slowly
converging Fourier series by integrating (B4) using
${\rm div} {\bf Q} = {\rm div} {\bf Q}_A = 0$ as in
Ref.~\onlinecite{15}. But it is more convenient to take
${\bf Q}_A$ from the exact relation
  \begin{eqnarray}    
  {\bf Q}_A({\bf r}) = {\nabla \omega_A \times {\bf\hat z}
  \over 2\, \kappa\, \omega_A } \,,
  \end{eqnarray}
where $\omega_A(x,y)$ is the Abrikosov $B_{c2}$ solution given
by the rapidly converging series (B1) with coefficients
\cite{30,31}
  \begin{eqnarray}    
    a_{\bf K}^A = -(-1)^{m+mn+n} \exp[ -K_{mn}^2 S/(8\pi)]
  \end{eqnarray}
for general lattice symmetry, and
$a_{\bf K}^A = -(-1)^{\nu^2} \exp(-\pi \nu^2 /\sqrt 3)$
($\nu^2 = m^2 +mn +n^2$) for the triangular lattice. This $\omega_A$
is normalized to $\langle \omega_A(x,y) \rangle =1$, which means
that $\sum'_{\bf K} a_{\bf K}^A = 1$ for any lattice symmetry.
Another strange property of the Abrikosov solution (B6) is that
$(\nabla\omega_A /\omega_A)^2 - \nabla^2 \omega_A /\omega_A =
 4\pi /S = {\rm const}$, although both terms diverge at the
vortex positions; this relation follows from (B4) and (B5)
using $ B = \Phi_0/S = 2\pi/(\kappa S)$. The useful
formula (B5) may be proven via the complex
$B_{c2}$ solution $\psi_A(x,y)$; it means that near $B_{c2}$
the third and fourth term in $F$, Eq.~(2), are identical.

  Approximate solutions $\omega({\bf r})$ and $B({\bf r})$ may be
computed by using a finite number of Fourier coefficients
$a_{\bf K }$ and $b_{\bf K}$ and minimizing the free energy
$F(B, \kappa, a_{\bf K }, b_{\bf K})$ with respect to these
coefficients \cite{15}. However, a much faster and more accurate
solution method \cite{17} is to iterate the two GL equations
$\delta F /\delta \omega =0$ and $\delta F /\delta {\bf Q} =0$
written in appropriate form. Namely, the iteration is stable
and converges rapidly if one isolates a term
$(-\nabla^2 + {\rm const}) (\omega, \, {\bf Q})$ on the
l.h.s.\ and puts the remaining terms to the r.h.s.\ as a kind
of ``inhomogeneity'' of such London-like equations, e.g.,
  \begin{eqnarray}    
  (-\nabla^2 +2\kappa^2)\, \omega ~&=&~ 2\kappa^2 (2 \omega
  -\omega^2 -\omega Q^2 -g) \,, ~\\
  (-\nabla^2 +\bar \omega)\, {\bf Q}_b ~&=&~ -\omega{\bf Q}_A
   -(\omega -\bar \omega){\bf Q}_b \,,
  \end{eqnarray}
with the abbreviations
$g({\bf r})=(\nabla\omega)^2 /(4\kappa^2 \omega)$,
${\bf Q}_b = {\bf Q - Q}_A$,
$\nabla\times {\bf Q}_b =B({\bf r}) - B$,
and $\bar\omega =\langle \omega \rangle =\sum'_{\bf K} a_{\bf K}$.
Equations (B7),  (B8) introduced some ``penetration depths''
$(2\kappa^2)^{-1/2}  = \xi /\sqrt2$  and
$\bar \omega^{-1/2} = \lambda/ \bar\omega^{1/2}$ (in real units),
which stabilize the convergence of the iteration. Acting on the
Fourier series $\omega$ (B1) and ${\bf Q}_b$ (B3) the
Laplacian $\nabla^2$ yields a factor $-K^2$; this facilitates
the inversion of (B7) and (B8). Using the orthonormality
  \begin{eqnarray}   
  2\, \langle \cos {\bf K r} \cos {\bf K' r} \rangle =
                                  \delta_{\bf K K'}
  \end{eqnarray}
(for ${\bf K} \ne 0$) one obtains from (B1), (B2)
$a_{\bf K} = -2\langle \omega({\bf r}) \cos{\bf Kr} \rangle$ and
$b_{\bf K} =  2\langle      B({\bf r}) \cos{\bf Kr} \rangle$.
The convergence of the iteration is considerably improved by
adding a third equation which minimizes $F$, Eq.~(2), with
respect to the amplitude of $\omega$, i.e.,
$\partial F / \partial \bar \omega = 0$. This step gives the
largest decrease of $F$. The resulting three iteration equations
for the parameters $a_{\bf K}$ and $b_{\bf K}$ then read \cite{17}
  \begin{eqnarray}    
  a_{\bf K} &:=& {4\kappa^2 \langle (\omega^2 +\omega Q^2 -2\omega
    + g ) \cos{\bf K r} \rangle \over K^2 + 2\kappa^2 } \,,
                                                      \\
  a_{\bf K} &:=& a_{\bf K} \cdot \langle \omega -\omega Q^2 -g
    \rangle ~/~ \langle \omega^2 \rangle\,,
                                                       \\
  b_{\bf K} &:=& {-2 \langle [(\omega -\bar\omega)B({\bf r})
    + p\, ] \cos{\bf K r} \rangle \over K^2 + \bar \omega } \,,
  \end{eqnarray}
with $p = (\nabla \omega \times {\bf Q}) {\bf\hat z} =
 Q_x \partial\omega /\partial y - Q_y \partial\omega /\partial x$
and $g=(\nabla\omega)^2 /(4\kappa^2 \omega)=(\nabla f)^2/\kappa^2$
as above.

  The solutions $\omega({\bf r})$, ${\bf B(r)}$, and ${\bf Q(r)}$
are obtained by starting, e.g., with
$a_{\bf K} = (1-b)\, a_{\bf K}^A $ and $b_{\bf K} = 0$
and then iterating the three equations (B10), (B11), (B12)
by turns until the coefficients do not change any more. After
typically 25 such triple steps, the solution stays constant
to all 15 digits and the GL equations are exactly satisfied.
Since all terms in (B10) - (B12) are smooth periodic
functions of ${\bf r}$, high accuracy is achieved by using
a regular spatial 2D grid, e.g.,
$x_i = (i- 1/2)x_1 /N_x$    ($i=1 \dots N_x$) and
$y_j = (j- 1/2)y_2 /(2N_y)$ ($j=1 \dots N_y$,
$2N_y \approx N_x y_2 /x_1$) with constant weights
$x_1/N_x$ and $y_2/(2N_y)$. These $N=N_x N_y$ = 100 to 5000
grid points fill the rectangular basic area $0\le x \le x_1$,
$0\le y \le y_2/2$, which is valid for any unit cell with the
shape of a parallelogram. Spatial averaging $\langle ...\rangle$
then just means summing $N$ terms and dividing by $N$.

  Best accuracy is achieved by considering all ${\bf K}_{mn}$
vectors within a half circle $|{\bf K}_{mn}| \le K_{max}$,
with $K_{max}^2 \approx 20 N/S$ chosen such that the number
of the ${\bf K}_{mn}$ is slightly less than the number $N$
of grid points.
 The high precision of this method may be checked with the
identity $B(x,y)/B_{c2} = 1 -\omega(x,y)$, which is valid at
$\kappa = 1/\sqrt 2$ for all $b$. This relation is confirmed
with an error $<10^{-9}$.
The equilibrium field $H$ or reversible magnetization
$M = B -H$ is computed from Doria's virial theorem, Eq.~(3).
\\

 \section{London theory}

  The modified London equation for a lattice of straight
vortex lines at regular positions ${\bf R = R}_{mn}$
(App.~B) is
  \begin{eqnarray}   
  (1- \lambda^2 \nabla^2)\, B(x,y) =
  \Phi_0 \sum_{\bf R} \delta({\bf r - R}_{mn}) \,,
  \end{eqnarray}
where $\delta(x,y)$ is the 2D delta function.
The solution for the magnetic field of one isolated vortex
at ${\bf R}=0$ is
  \begin{eqnarray}   
  B_v(r) = (\Phi_0 / 2\pi \lambda^2)\, K_0(r/\lambda) \,.
  \end{eqnarray}
The modified Bessel function
  \begin{eqnarray}   
  K_0(r/\lambda) = { \int {d^2 k \over 2\pi}\,
    {\cos {\bf k r}  \over \lambda^{-2} + k^2 } }
  \end{eqnarray}
has the derivative $K_0(x)' = -K_1(x)$ with the limits
$K_0(x\!\ll\!1) \approx -\ln x$,
$K_1(x\!\ll\!1) \approx 1/x$, and for $x\!\gg\!1$: \cite{32}
  \begin{eqnarray}   
  K_0(x)\! &\approx& \!\sqrt{ {\pi \over 2x} }\, e^{-x}
   \!\left(1 -
   {1\over 8x} +{9\over128x^2}-{225\over3972x^3}\right),~~~
     \nonumber \\
  K_1(x)\! &\approx& \!\sqrt{ {\pi \over 2x} }\, e^{-x}
   \!\left(1 +
   {3\over 8x}-{15\over128x^2}+{315\over3972x^3}\right) .~~
  \end{eqnarray}
For a periodic FLL one obtains the Fourier series $B(x,y)$,
Eq.~(16), which may also be written as a sum over isolated
vortex fields, $ B(x,y)= \sum_{\bf R} B_v({\bf r - R})$.
Similarly, the free energy of the FLL may be written as a
sum of vortex self energies ($\Phi_0 H_{c1}$ per unit length)
plus a double sum over all interactions between two vortices.
The average energy density $F$, Eq.(17), then reads
  \begin{eqnarray}   
  F = B H_{c1} +{B\Phi_0 \over 4\pi\lambda^2\mu_0 }
  \sum_{\bf R} K_0( R / \lambda) \,.
  \end{eqnarray}
For the triangular vortex lattice we write $R/\lambda=\nu c$
with $c =a/\lambda = (4\pi/\sqrt3)^{1/2} (b\kappa^2)^{-1/2}$
($a$ = vortex spacing) and
$\nu^2 = m^2 +mn +n^2 =$ 1, 3, 4, 7, 7, 9, $\dots$ .
Taking the derivative $H = \partial F / \partial B$ one obtains
for $h=H/H_{c2}$ with $h_{c1} = H_{c1}/H_{c2}$:
  \begin{eqnarray}   
  h = h_{c1} +{3 \over \kappa^2}  \sum_\nu \left[ K_0(\nu c) +
     {\nu c\over 2} K_1(\nu c) \right] .
  \end{eqnarray}
Here the sum is over $\nu = 1, \sqrt3, 2, \dots$ , i.e.\ the
number of six flux lines per shell is already accounted for.
Equation (C6) is still exact. It works for $b \ll 1$
(i.e. for nonoverlapping vortex cores) and for $\kappa > 1.4$
(i.e. when the long-range interaction of vortices is purely
magnetic \cite{11,33}). With the expansions (C4) one obtains
for $x=\nu c \gg 1$:
  \begin{eqnarray}   
  h \approx h_{c1} + {3 \sqrt\pi \over 2 \kappa^2} \sum_\nu
  e^{-x} \sqrt{x} \left[1+{19\over8x} -{47\over128x^2} \right].
  \end{eqnarray}
At very small $b$, namely for $c = a/\lambda \gg 1$, the sum
may be restricted to the nearest neighbor shell,
i.e. to the first term, $\nu=1$, yielding
  \begin{eqnarray}   
  h \approx h_{c1} + {3 \sqrt{\pi c} \over 2 \kappa^2}
  e^{-c}  \left[1+{19\over8c} -{47\over128c^2} \right].
  \end{eqnarray}



\end{document}